\documentclass[aps,prl,twocolumn,preprintnumbers,showpacs,superscriptaddress,nofootinbib,floatfix,10pt]{revtex4-1}
\pdfoutput=1
\usepackage{textcomp}
\usepackage[english]{babel}
\usepackage{amsmath,amssymb,amsbsy,booktabs}
\usepackage{bm}
\usepackage{color}
\usepackage{array}
\usepackage{amstext}
\usepackage{graphicx}
\usepackage{amsfonts}
\usepackage{bm}
\usepackage{dcolumn}
\usepackage{rotating}
\usepackage{epstopdf}
\usepackage{esint}
\usepackage{url}
\usepackage{slashed}
\usepackage{fontawesome5}
\usepackage[colorlinks=true,
linkcolor=blue,
filecolor=blue,
anchorcolor=blue,
urlcolor=blue,
citecolor=blue
]{hyperref}

\usepackage[T1]{fontenc} 
\usepackage[utf8]{inputenc} 
\usepackage{times}
\usepackage{float}
\usepackage{subfigure}

\definecolor{color_git}{rgb}{0.098, 0.160, 0.345}
\newcommand{\gitlink}{\href{https://github.com/Shao-Ping-Li/Scattering_vs_Forbidden_Decay}{\textsc{g}it\textsc{h}ub {\large\color{color_git}\faGithub}}} 

\allowdisplaybreaks[0]

\begin{document}
	
	\title{\bf Scattering versus Forbidden  Decay in Dark Matter Freeze-in} 
	
	\author{Shao-Ping Li}
	\email{spli@ihep.ac.cn}
	\affiliation{Institute of High Energy Physics, Chinese Academy of Sciences, Beijing 100049, China}

	\begin{abstract}
It is generically believed that the  two-body scattering is  suppressed   by higher-order weak couplings with respect to the two-body decay. We show that this does not always hold when a heavy particle is produced by  forbidden decay  in a thermal plasma,  where the scattering   shares  the same order of couplings   with the  decay. We find that there is a simple and close relation between  the forbidden decay and  the same-order scattering.  To illustrate this point, we consider  freeze-in production of   heavy dark matter via a light scalar mediator. We point  out that, when the Boltzmann (quantum) statistics is used,  the forbidden decay   can contribute to the   dark matter relic density  at 5$\%$-24$\%$ (10$\%$-$39\%$)  with a weak thermal coupling,  while   the  contribution from the scattering channel can be several orders of magnitude  larger than   from the forbidden decay  if the  thermal coupling is much smaller. 
Such a relative effect between the scattering and the forbidden decay  could  also exist  in  other plasma-induced processes, such as the purely thermal generation of the right-handed neutrino dark matter, or of  the  lepton asymmetry in leptogenesis.    
	\end{abstract}
	
	
	\maketitle

\textbf{Introduction}.---
In many theories beyond the Standard Model (SM) of particle physics, a heavy  species can be usually produced by a light particle in a thermal plasma. This kind of production,  which is kinematically forbidden in vacuum  but opened at finite temperatures due to plasma effects, has been studied in a wide range of phenomena, such as the  dark matter (DM) production~\cite{Rychkov:2007uq,Strumia:2010aa,Drewes:2015eoa,Baker:2017zwx,Dvorkin:2019zdi,Darme:2019wpd,Biondini:2020ric,Konar:2021oye}, the production of    neutrinos from plasmon decay in stellar cooling~\cite{Bernstein:1963qh,Braaten:1993jw,Raffelt1996,Yakovlev:2000jp,Hardy:2016kme}, and   the thermally induced baryon asymmetry in the early universe~\cite{Giudice:2003jh,Garny:2009qn,Garny:2009rv,Kiessig:2010pr,Garny:2010nj,Kiessig:2011fw,Kiessig:2011ga,Garbrecht:2012qv,Hambye:2016sby,Hambye:2017elz,Li:2020ner,Li:2021tlv}. 

 In the scenarios of   forbidden decay,  the  two-body scattering mediated by the light particle can also be significant and even dominate the production.
A known example is the  neutrino chirality-flipping process $\nu_L\to \nu_R$ in the relativistic QED plasma,  where the contribution from the $t$-channel scattering  $e+\nu_L\to e +\nu_R$    was  found to be much  larger than   from the plasmon decay $\gamma^*\to \bar\nu_L + \nu_R$,  since the latter is suppressed by a  higher-order electromagnetic coupling $\alpha_{\rm EM}$~\cite{Fukugita:1987uy,Elmfors:1997tt,Ayala:1999xn,Li:2022dkc}. A similar effect is also found recently in the electron  chirality-flipping  process~\cite{Boyarsky:2020cyk,Boyarsky:2020ani}.
For   a nonthermal DM  produced via the freeze-in paradigm~\cite{McDonald:2001vt,Kusenko:2006rh,Petraki:2007gq,Hall:2009bx,Bernal:2017kxu},   it has been shown that the forbidden two-body  decay can be the dominant mechanism (see e.g. Refs.~\cite{Rychkov:2007uq,Dvorkin:2019zdi,Chang:2019xva}) and in some cases be the unique channel to account for the DM relic density~\cite{Darme:2019wpd,Konar:2021oye}. 

\begin{figure}[t]
	\centering
	\includegraphics[scale=1.1]{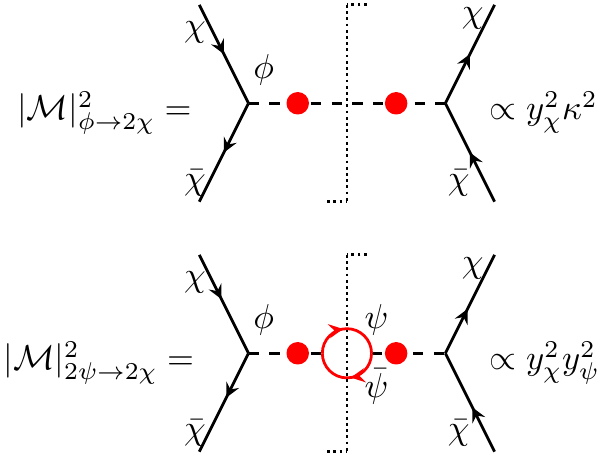}
	\caption{The  scattering channel  $\bar\psi+\psi\to \bar\chi+\chi$ associated with the forbidden decay $\phi\to \bar\chi+\chi$ at the same   order of coupling prefactor $ y_\psi^2 y_\chi^2$, where $\kappa\sim y_\psi$ is generically expected when the thermal correction (red blob) to $\phi$  dominantly arises from a self-energy topology  similar to the red one.
	}
	\label{fg:FScat}
\end{figure}

The  two-body scattering is generically expected to be  suppressed   by higher-order weak couplings,  which results in  the two-body decay being the dominant channel for most situations.  However, when  the      decay channel is   a purely plasma-induced effect, the two-body scattering  associated with the very forbidden decay can carry the same order of coupling constants.
  To see this, we show  an example in  Fig.~\ref{fg:FScat} with the  scalar forbidden decay to fermions  via the Yukawa interaction $y_\chi\bar\chi \chi \phi$.  For a vacuum mass condition $m_\phi<2m_\chi$, the scalar  decay $\phi \to \bar\chi+\chi$ is   kinematically forbidden in vacuum but opened at   temperatures above some critical point $T_{\rm c}=2m_\chi/\kappa$ as the light  scalar  $\phi$ acquires temperature-dependent thermal mass $m_{\phi}(T)\equiv \kappa T$ from, e.g.,  the Yukawa interaction $y_\psi \bar \psi\psi\phi$. Here $\kappa$  characterizes the correction factor   from  the thermal plasma, which is    encoded in the red blob of Fig.~\ref{fg:FScat}. Since  a nonzero $\kappa$ is induced by the resummed self-energy in the red blob, it points out that the   scattering $\bar\psi+\psi\to\bar\chi+\chi$   mediated by the light scalar 
also exists when the  forbidden decay is opened. This can be seen by cutting the red blob in the forbidden decay diagram such that the loop particles go on-shell while the scalar becomes off-shell. With such a cut, the scattering channel is said to be hinted from the forbidden decay diagram.

As will be derived in this paper,  both the scattering and forbidden decay rates can carry the same order of coupling prefactor $\propto y_\chi^2y_\psi^2$. It  differs from the usual vacuum situations  where the scattering carries higher-order weak couplings, and  also from   the chirality-flipping processes where the scattering channel  carries a lower-order electromagnetic coupling as mentioned above.  Without the suppression (enhancement) of higher (lower)-order weak couplings in the scattering channel, it could  be nontrivial  to see the relative effect  of the   plasma-induced  decay and the  scattering. In particular,  whenever   nonthermal DM production with a light thermal mediator   is concerned,  it would be tempting  to know the portion from mediator forbidden decay  when  the conventional scattering via a light mediator is considered~\cite{Chu:2011be}. On the other hand, whenever the   forbidden decay from a  light mediator can account for the DM relic density~\cite{Rychkov:2007uq,Dvorkin:2019zdi,Chang:2019xva,Darme:2019wpd,Konar:2021oye}, it would be necessary to check if the scattering effect is indeed suppressed. 

 In this paper, we consider a close relation between  the renormalizable forbidden decay and the associated scattering channel in freeze-in DM production via a light mediator. 
 To illustrate the key formulation for comparing  the scattering with the forbidden decay,  
we   consider in the remainder of this paper  a light scalar mediator shown  in Fig.~\ref{fg:FScat}, where the nonthermal fermion $\chi$ is a DM candidate having  a direct freeze-in channel from the scalar forbidden decay $\phi\to \bar\chi+\chi$.  
The close relation to be shown can be simply characterized by the dominant coupling that helps the mediator   to equilibrate with the thermal plasma.  Moreover, the simple relation allows us to estimate the relative contribution of forbidden decay and scattering in  producing the observed DM relic density.  

As will be  shown below,   the scattering effect   can  dramatically modify the forbidden decay scenarios of DM production at finite temperatures.  In particular, the ratio of the relic density from the  scattering  to that from the forbidden decay  has a simple scaling $\sim 1/y_\psi$ in the  weak-coupling  limit $y_\psi<1$. It implies that   the contribution from the forbidden decay   can only become significant for a large thermal coupling, and if not, the scattering contribution will      be  orders of magnitude  larger than  from the pure decay channel.

 We expect that the close relation can  also    exist in a wide range of  scenarios, such as the  millicharged DM~\cite{Davidson:2000hf,Chang:2018rso} generated from the plasmon decay~\cite{Dvorkin:2019zdi}, the right-handed neutrino DM~\cite{Drewes:2016upu,Boyarsky:2018tvu} from a thermal scalar decay~\cite{Drewes:2015eoa},  or even  the nonthermal DM production from a  hidden thermal plasma~\cite{Feng:2008mu,Berlin:2016gtr}. We further expect that it  could      modify the pattern of leptogenesis  when the out-of-equilibrium generation of lepton  asymmetries results from   forbidden decay in the  early universe~\cite{Giudice:2003jh,Hambye:2016sby,Hambye:2017elz,Li:2020ner,Li:2021tlv}.
The investigation presented here  complements the widely studied plasma-induced  effects at finite temperatures   where  the two-body decay and the scattering generically carry different  powers of coupling  prefactors. 

 \begin{figure*}[t]
 	\centering
 	\includegraphics[scale=0.6]{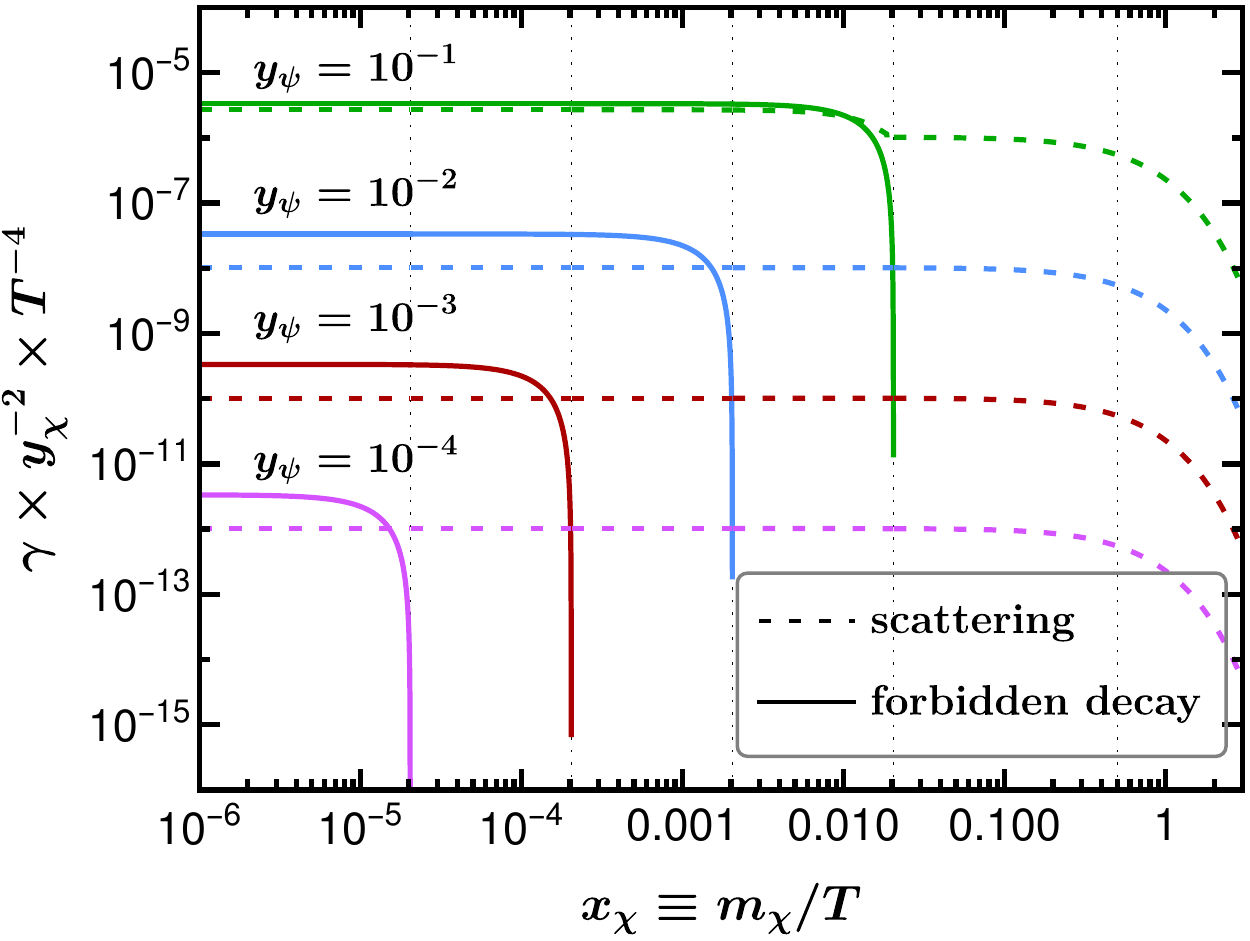}\qquad
 	\includegraphics[scale=0.6]{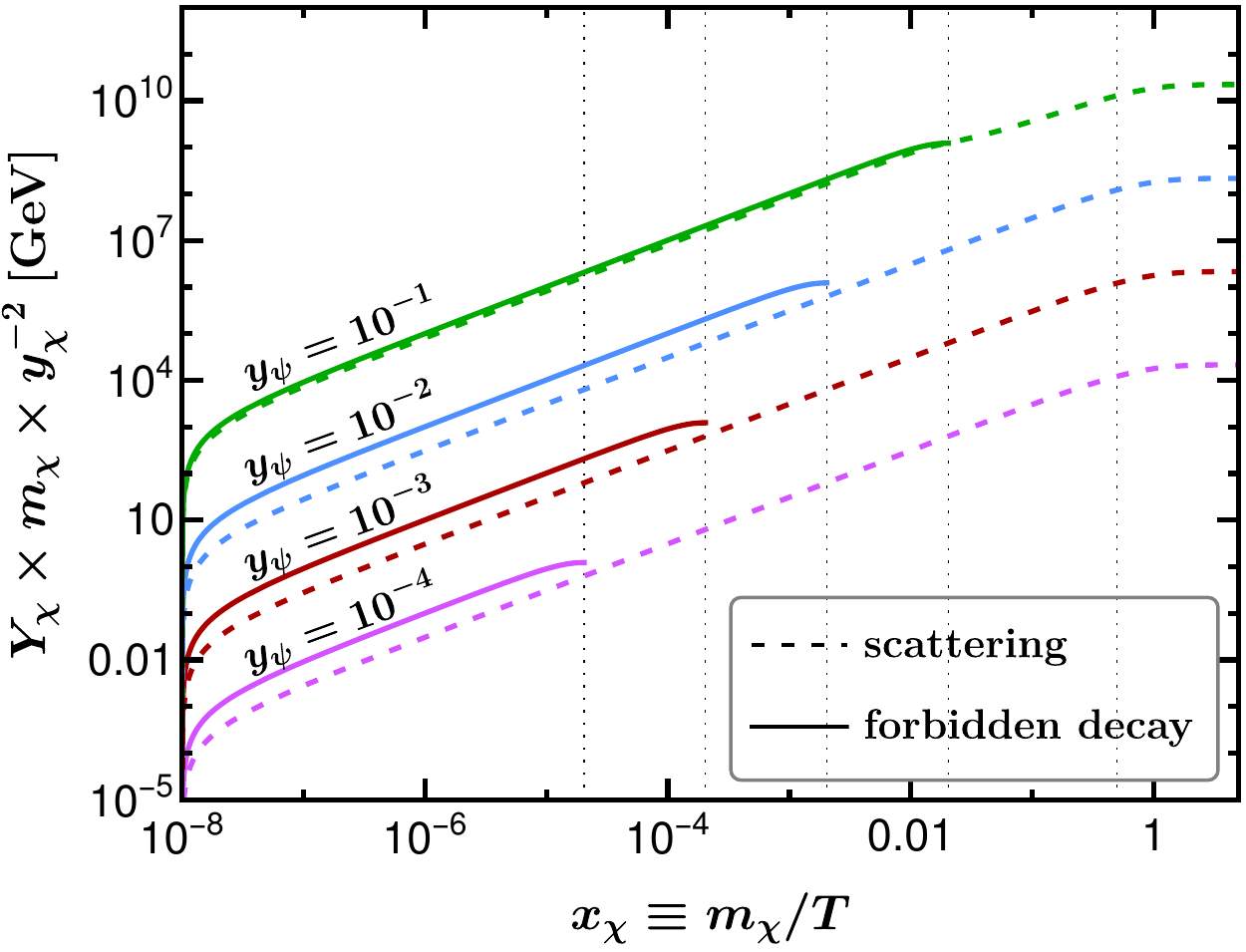}
 	\caption{ Left: comparison of collision rates between the forbidden decay and the off-shell scattering.  The collision rates are normalized to the squared DM coupling $y_\chi^2$ and the quartic temperature $T^{4}$. Right: comparison of DM yields $Y_\chi$  by factoring out the dependence on the DM mass $m_\chi$ and the DM coupling $y_\chi$. The code for obtaining the figures is publicly available at \gitlink.
 		\label{fg:resonant}  
 	}
 \end{figure*}
 
\textbf{Relative rates of  forbidden decay and scattering}.---
Let us first point out that the scattering effect at high temperatures  could be already comparable to the forbidden decay rate.    The   heavy DM $\chi$ can be produced by a light thermal scalar $\phi$ which establishes thermal equilibrium with a light fermion $\psi$ in the plasma. The relevant interaction is characterized by
	\begin{align}\label{eq:lag}
		\mathcal{L}=y_\chi \bar \chi \chi \phi+y_\psi \bar\psi \psi \phi.
	\end{align}
We   consider the situation where   thermal particles $\phi,\psi$ are  much lighter than the nonthermal $\chi$ at zero temperature so that the  nonrelativistic   $\bar\psi\psi$  annihilation  to $\chi$ is  kinematically forbidden  in vacuum.  In practice, the subsequent calculations are approximately obtained by taking $m_{\phi,\psi}(T=0)=0$. It should be mentioned, however, when the vacuum masses of the thermal particles are compatible with $m_\chi$, significant corrections could arise.  
For simplicity, we further assume that the dominant thermal correction to   $\phi$ can be well encapsulated by the $\bar\psi \psi \phi$ interaction. Including other comparable corrections opens additional   scattering channels associated with the forbidden decay.  While we are not devoted to specific scenarios, it it noteworthy  that realistic models can already be constructed from~\eqref{eq:lag}. For example, the scalar is a SM singlet and couples to a light Majorana neutrino $\psi$ which can readily equilibrate with the SM plasma via neutrino oscillation~\cite{Dolgov:2003sg,Li:2022bpp}.

A special exception that the  scattering channel hinted from the forbidden decay diagram may be suppressed  is the $\lambda \phi^4$ theory, which generates $\kappa\propto \sqrt{\lambda}$ at leading order~\cite{Bellac2000}. The $\lambda \phi^4$ interaction can also induce a $\phi^3$ vertex $\propto \lambda v_\phi$ when $\phi$ develops a nonzero vacuum expectation value $v_\phi$. Then, besides the additional dependence on $v_\phi$, the collision rate from  the scattering $\phi+\phi\to \bar\chi+\chi$ has a higher-order $\lambda$  prefactor than from the forbidden decay $\phi\to \bar\chi+\chi$, and would be  suppressed by small $\lambda$.
Nevertheless, when $\lambda$ is   small,  the scattering  comparable with the forbidden decay can still be opened from e.g.   a   gauge  scalar-vector-vector   $B_\mu B^\mu \phi$  or a trilinear-scalar $\phi \Phi^2$ interaction. 
Therefore the  same-order scattering associated with the forbidden decay can be a generic result  when the dominant thermal correction   arises from a self-energy diagram similar to  the red bubble in Fig.~\ref{fg:FScat}.  

From~\eqref{eq:lag},   the  squared amplitude  in the forbidden decay $\phi\to\bar\chi+ \chi$ reads
\begin{align}\label{eq:damp}
	|\mathcal{M}|_{\phi\to 2\chi}^2=2y_\chi^2  \left(\kappa^2 T^2- 4m_\chi^2\right).
\end{align}
The Boltzmann  equation for the evolution of $\chi$ number density is given by
\begin{align}
	\dot{n}_\chi+3Hn_\chi =2\gamma_{\phi\to 2\chi},
\end{align}
where the Hubble parameter reads $H\approx 1.66  \sqrt{g_{\rho}}T^2/ M_{\rm Pl}$, with $M_{\rm Pl}=1.22\times 10^{19}$~GeV  the Planck mass  and  $g_{\rho}$ the effective degrees of freedom for the energy density. The factor of  2   results from  the $\chi$-pair production.
The collision rate from Eq.~\eqref{eq:damp} reads
\begin{align}
	\gamma_{\phi\to 2\chi}&=\frac{\kappa^3y_\chi^2 K_1(\kappa)}{16\pi^3}\left(1-\frac{4m_\chi^2}{\kappa^2 T^2}\right)^{3/2}T^4,
\end{align}
where $K_1(\kappa)$ is the modified Bessel function with $K_1(\kappa)\approx 1/\kappa$. Note that in obtaining the collision rate in  the Boltzmann equation, we  apply the  Boltzmann  distribution $f=e^{-E/T}$ for the thermal particles and the Pauli-blocking effect from the  nonthermal DM  $\chi$ is neglected. To make a comparison between the Boltzmann approximation and the full quantum statistics for the thermal particles, we will perform the analysis of the full quantum statistics  whenever relevant in the subsequent discussions. For the moment, it suffices to take the Boltzmann  distribution as an approximation to analyze the relative effect of the scattering and forbidden decay channels.

The scattering production for the nonthermal DM $\chi$  occurs through the  $s$-channel  $\bar\psi+\psi\to \bar\chi+\chi$. With the usual treatment in vacuum, the cross section without the spin average of $\psi$  is  simply given by
\begin{align}\label{eq:CShighQ}
	\sigma_{2\psi\to2\chi}=\frac{y_\chi^2 y_\psi^2}{4\pi s}\left(1-\frac{4m_\chi^2}{s}\right)^{3/2}.
\end{align}
The resulting collision rate  with the Boltzmann statistics  then reads~\cite{Gondolo:1990dk}
\begin{align}\label{eq:gammascat}
	\gamma_{2\psi\to2\chi}&\approx \frac{ T}{32 \pi^4} \int_{4m_\chi^2}^\infty ds	\sigma_{2\psi\to 2\chi} s^{3/2}K_{1}(\sqrt{s}/T).
\end{align}
It can be seen that, the scattering rate  in this vacuum treatment can already be comparable to the forbidden decay rate if $\kappa$ is at $\mathcal{O}(y_\psi)$.
Explicitly, using $\kappa=y_\psi/\sqrt{6}$ to be derived below, we have  the approximate  relation 
\begin{align}\label{eq:dsrate}
\frac{\gamma_{2\psi\to2\chi}}{ \gamma_{\phi\to2\chi}}\approx 0.3
\end{align}
in the high-temperature limit $m_\chi/T\ll 1$.  Note that, however, if the full quantum statistics is used, i.e., Fermi-Dirac distribution for the thermal fermion $\psi$ and Bose-Einstein distribution for the thermal scalar $\phi$, the ratio  in Eq.~\eqref{eq:dsrate} becomes 0.13. This suppression arises from the Pauli-blocking effects for $\psi$ and Bose enhancement for $\phi$.

Thus far the cross section is only computed in the limit of $s\gg m_\phi^2(T)$, where the effect near the pole $s=m_\phi^2(T)$ is not taken into account properly. Since the  cross section may be   enhanced near the pole and both $\gamma_{\phi\to 2\chi}$ and  $\gamma_{2\psi\to 2\chi}$ have the same prefactor dependence,  the effect from such an $s$-channel enhancement   could further increase the ratio   given by Eq.~\eqref{eq:dsrate}.
The resonant enhancement appears  when the momentum transfer is at $\mathcal{O}(\kappa T)$.   This soft-scattering transfer can  come either from the soft   $\bar\psi \psi$ pair with momenta at $\mathcal{O}(\kappa T)$, or from the collinear   $\bar\psi \psi$ pair with hard  momenta at $\mathcal{O}(T)$  but with a small  angle at  $\mathcal{O}(\kappa)$ between the $\bar\psi\psi$ momenta~\cite{Arnold:2002zm,Li:2023ewv}. Under the perturbative Hard-Thermal-Loop  (HTL) technique~\cite{Braaten:1989mz,Frenkel:1989br,Braaten:1991gm} (see also e.g. Refs.~\cite{Carrington:1997sq,Bellac2000}),      the thermal correction  to  $\psi$  for   hard $\bar\psi\psi$ pair  is  of  higher order, while for  soft $\bar\psi\psi$ pair, both the thermal correction  to $\psi$ and the resummed $\bar\psi \psi \phi$ vertex should be included to obtain a consistent result at leading order. Here we consider the hard $\bar\psi\psi$ pair since a thermal relativistic particle has an averaged momentum at $\mathcal{O}(T)$.

Following the effective treatment in Ref.~\cite{Arnold:2002zm},  we   compute the cross section by  
including the leading-order thermal correction in the internal $\phi$ propagator and treating the external hard $\bar\psi \psi$ pair effectively massless. The cross section reads
\begin{align}\label{eq:fullscat}
	\sigma_{2\psi\to 2\chi}=\frac{y_\chi^2y_\psi^2}{4\pi\sqrt{s}}  \frac{(s-4m_\chi^2)^{3/2}}{[s-\text{Re}\Pi_R^\phi]^2+[	\text{Im}\Pi_R^\phi]^2},
\end{align}
where $\Pi_R^\phi$ is the resummed retarded self-energy amplitude of $\phi$. 
In the real-time formalism of thermal field theory~\cite{Landsman:1986uw,Ghiglieri:2020dpq},  the real part of  $\Pi_R^\phi$ is given by
\begin{align}\label{eq:PiR-phi}
	\text{Re}\Pi_R^\phi&=\frac{y_\psi^2}{\pi^3} \int d^4q f_\psi(\omega)  \frac{k.q}{(k+q)^2}\delta(q^2),
\end{align} 
where $f_\psi(\omega)=(e^{\omega/T}+1)^{-1}$ is the Fermi-Dirac distribution function for $\psi$ with $\omega\equiv |q_0|$ 
and $k$ is the 4-momentum of  $\phi$ with $s=k^2$.  
In the HTL approximation,  $k^2/|\vec q|^2\sim \mathcal{O}(y_\psi^2)$ is of higher order. We  neglect   these high-order terms  in the integral and obtain
$	\text{Re}\Pi_R^\phi \approx y_\psi^2T^2/6$.
The dispersion relation of the  thermal scalar $\phi$ is determined by the pole $k^2-\text{Re}\Pi_R^\phi=0$, leading to    $\kappa=y_\psi/\sqrt{6}$. 
On the other hand, the imaginary self-energy amplitude is   given by   
\begin{align}
	\text{Im}\Pi_R^\phi=-\frac{y_\psi^2 k^2}{4\pi^2}\int d^4 q  [1-2f_\psi(\omega)]\delta_{k+q}\delta_{q},
\end{align}
where the two Dirac $\delta$-functions $\delta_{k+q}\equiv \delta[(k+q)^2]$ and $\delta_q\equiv \delta(q^2)$  dictate that the  loop particles  go on-shell. 

 Since the scattering has an $s$-channel resonance, the collision rate in the Boltzmann equation should be calculated without double counting~\cite{Arnold:2002zm}. There are several  methods to  remove the double counting~\cite{Cline:1993bd,Giudice:2003jh,Pilaftsis:2003gt,Cline:2017qpe,Belanger:2018ccd,DeRomeri:2020wng,Ala-Mattinen:2022nuj,Bringmann:2021sth,Li:2023ewv}.
 Here, we follow Refs.~\cite{Cline:1993bd,Cline:2017qpe,Ala-Mattinen:2022nuj} with a real-intermediate-state subtraction by splitting  the Breit-Wigner form of the scalar propagator as
 \begin{align}
 	iG_\phi(p^2)&=\frac{i}{p^2-m_\phi^2+i m_\phi \Gamma_\phi}
 \nonumber	\\[0.2cm]
 	&=\frac{i(p^2-m_\phi^2)}{(p^2-m_\phi^2)^2+m_\phi^2 \Gamma_\phi^2}+\frac{m_\phi \Gamma_\phi}{(p^2-m_\phi^2)^2+m_\phi^2 \Gamma_\phi^2}
 	\nonumber	\\[0.2cm]
 	&\equiv i G_{\phi,\rm off}(p^2)+G_{\phi,\rm on}(p^2),
 \end{align}
where $\Gamma_\phi$ is the decay width of the thermal scalar. In practice,  we use the thermal scalar mass, $m_\phi^2=\text{Re}\Pi_R^\phi$, and take $\Gamma_\phi=\text{Im}\Pi_R^\phi(k^2=m_\phi^2)/m_\phi$  to  estimate  the damping rate of the thermal scalar from the simplified Lagrangian~\eqref{eq:lag}.  For off-shell scattering, the scalar propagator is given by the off-shell term $G_{\phi,\rm off}$. The off-shell part  from Eq.~\eqref{eq:fullscat} is given by
\begin{align}\label{eq:offscat}
	\sigma_{2\psi\to 2\chi,\rm off}=\frac{y_\chi^2y_\psi^2 \beta}{4\pi}\frac{s(s-m_\phi^2)^2}{[(s-m_\phi^2)^2+m_\phi^2\Gamma_\phi^2]^2},
\end{align}
with $\beta\equiv (1-4m_\chi^2/s)^{3/2}$, which is then
 substituted  into  Eq.~\eqref{eq:gammascat} to obtain the collision rate in the Boltzmann approximation. 

The comparison between the  forbidden decay and  the  scattering   is shown in Fig.~\ref{fg:resonant}.   In the Boltzmann approximation, 
the ratio given in Eq.~\eqref{eq:dsrate}  at $T\gg T_c$ is kept for  $y_\psi=10^{-4}-10^{-2}$ but  enhanced to be $\gamma_{2\psi\to2\chi,\rm off}\approx 0.81\gamma_{\phi\to2\chi}$ for $y_\psi=0.1$. Therefore, the relation in Eq.~\eqref{eq:dsrate} can be lifted up by a factor of $\mathcal{O}(1)$ for a large thermal coupling $y_\psi$. With   a large $y_\psi$,   the thermal  correction  included in the propagator   can partially compensate for the suppression of additional phase-space factors in the two-body scattering.  
When the full quantum statistics is used for the initial thermal particles, we found that $\gamma_{2\psi\to2\chi,\rm off}\approx 0.13\gamma_{\phi\to2\chi}$ for     $y_\psi=10^{-4}-10^{-1}$, which is the same as in the vacuum case as discussed below Eq.~\eqref{eq:dsrate}.
It implies that the enhancement of the ratio $\gamma_{2\psi\to2\chi,\rm off}/\gamma_{\phi\to2\chi}$  due to large thermal corrections of the mediator propagator becomes  less significant when the Pauli-blocking  effects of the two thermal fermions and the Bose enhancement of the thermal scalar are accounted for in the collision rates.

When $T$ evolves down to  the  threshold point $T_{\rm c}=2m_\chi/\kappa$ (the vertical dotted lines),  the kinematic space for the forbidden decay tends to close, thereby exhibiting a sudden drop in the left panel of  Fig.~\ref{fg:resonant}. 
Nevertheless, the scattering continues  
 until $T$ drops below $m_\chi$,  after which the scattering rate will carry a Boltzmann suppression factor $e^{-m_\chi/T}$, as shown by the   drop of the collision curves. 
 
Besides a potential $\mathcal{O}(1)$  enhancement  near the resonance region, there is a  more important effect   after the decay channel closes. As seen in   the left panel of Fig.~\ref{fg:resonant}, there is a period of $\chi$ production from the pure   scattering channel  while  the duration  of the forbidden decay depends on the thermal coupling  $y_\psi$. For smaller $y_\psi$, the decay duration is shorter and   hence less $\chi$ production. This is explained by the fact that smaller $y_\psi$ dictates a higher threshold temperature $T_c$, consequently leading to a shorter duration of the forbidden decay in the early universe. This observation 
 implies that the contribution from the  scattering can be much larger than from  the forbidden decay   if the  pure scattering  lasts  sufficiently long in the expansion history of the universe. We show in the right panel  of Fig.~\ref{fg:resonant} a complementary plot for the DM yield $Y_\chi\equiv n_\chi/s_{\rm SM}$ as a function of  $x_\chi\equiv m_\chi/T$. It can be seen that the generation of $Y_\chi$ ends at the critical temperature (the vertical dotted lines) in  the forbidden decay channel but continues  below $T_c$ in the scattering channel. It points out clearly that the final abundance from  the forbidden decay channel can  become comparable to the scattering channel when  the thermal coupling $y_\psi$ becomes large.

\begin{figure*}[t]
	\centering
	\includegraphics[scale=0.6]{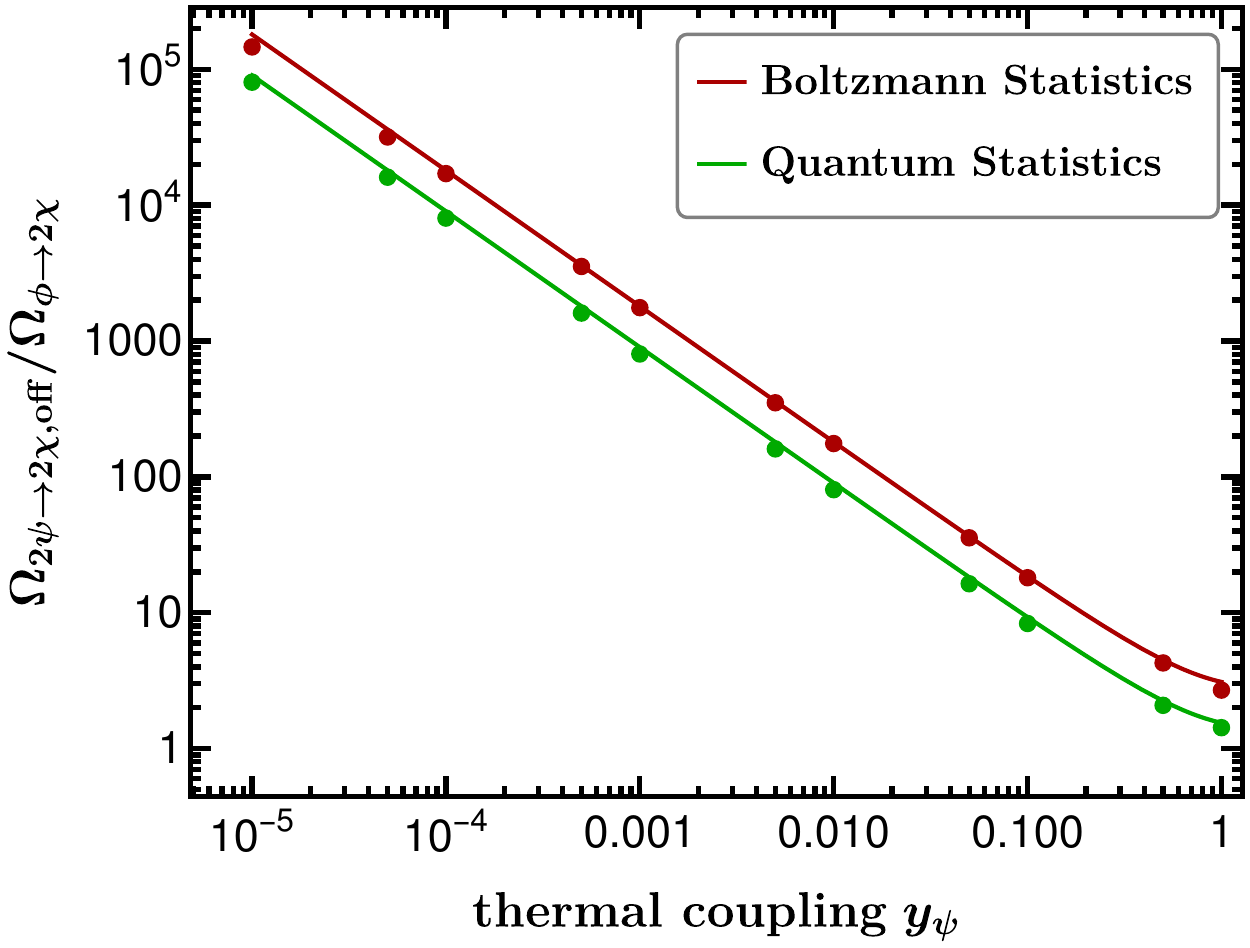}\qquad
	\includegraphics[scale=0.6]{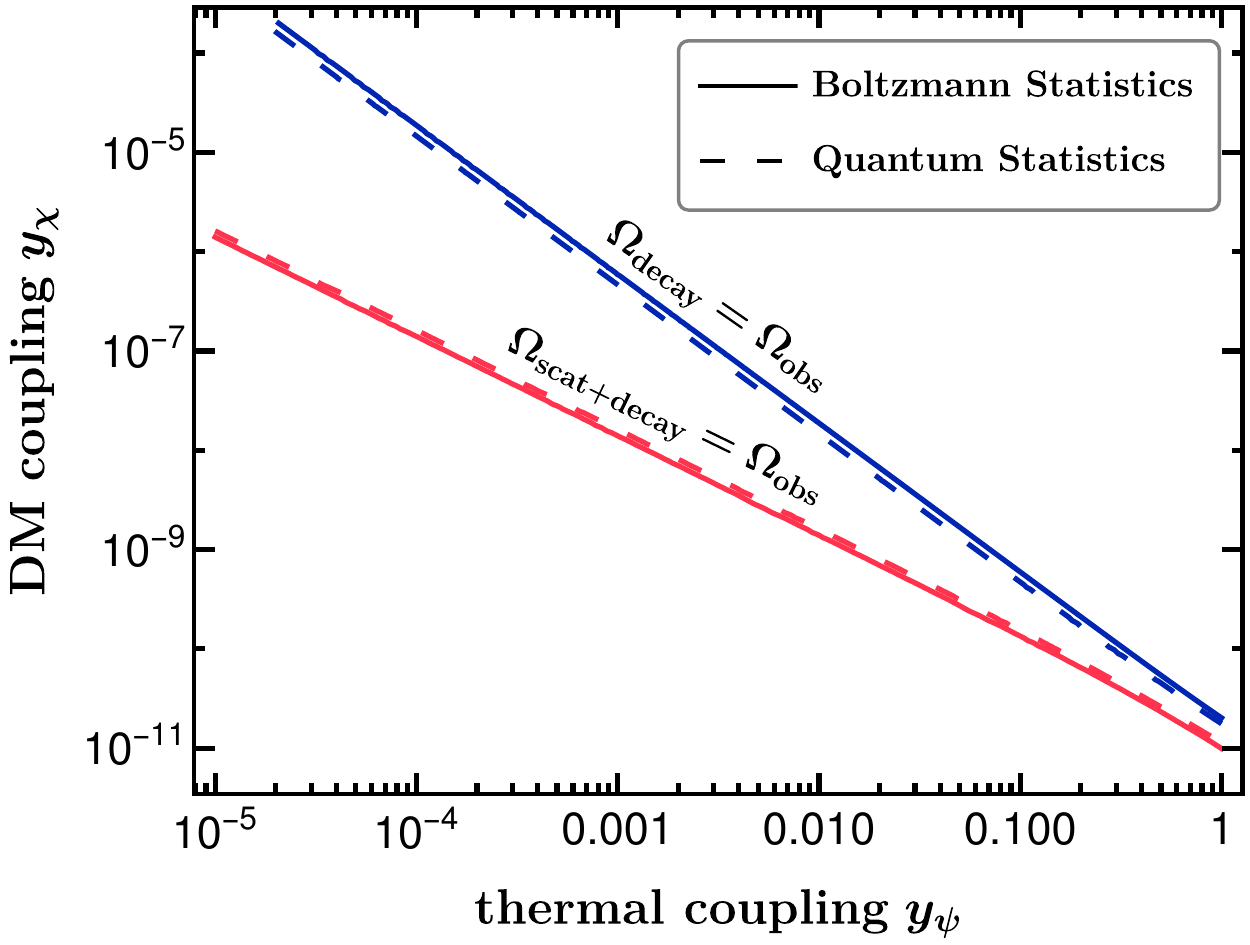}
	\caption{Left: the relic density ratio of off-shell scattering to forbidden decay. The selected points are obtained from full numerical results, while the solid lines are obtained from the fitted formula~\eqref{eq:OmegaR}. Right: the correlation between the DM coupling $y_\chi$ and the thermal coupling $y_\psi$ for the observed DM relic density obtained from the forbidden decay   only and the sum of  the forbidden decay and    scattering channels, respectively.   The results from the  Boltzmann approximation and the full quantum statistics are shown as a comparison.
		\label{fg:DMrelic}
	}
\end{figure*}
 
\textbf{Relative DM relic density}.---
To see the relative contribution of  the  scattering   and the forbidden decay to the DM relic density, we solve the  Boltzmann equation of  the DM number density yield  from 
\begin{align}\label{eq:Boltzmanneq}
Y_\chi \approx \int_{T_c}^\infty\frac{2\gamma_{\phi\to 2\chi}}{s_{\rm SM} H T}dT+\int_{0}^\infty\frac{2\gamma_{2\psi\to 2\chi,\rm off}}{s_{\rm SM} H T}dT,
\end{align}
where   $s_{\rm SM}=g_{s} 2\pi^2 T^3/ 45$ is the   SM entropy density with $g_{s}$ the effective degrees of freedom, and $\gamma_{2\psi\to 2\chi,\rm off}$ is the off-shell scattering rate.  We have used the symbol $\approx$ above  to highlight that the yield is approximately obtained in the limit of $m_{\phi,\psi}(T=0)=0$. The forbidden decay is closed at $T_c=2m_\chi/\kappa$, and  the   scattering essentially ends around $T\simeq m_\chi$ but using $T=0$ as the lower limit in Eq.~\eqref{eq:Boltzmanneq} does not cause significant   difference.  In the following, we will not distinguish  the small difference  between $g_{s}$ and $g_{\rho}$,  and simply set $g_{s}=g_{\rho}=106.75$  which becomes a good approximation if   the freeze-in temperature is well above the GeV scale~\cite{Husdal:2016haj}. 

 To compare the relic densities produced  from the forbidden decay and the    scattering, we integrate the temperature analytically for $\gamma_{\phi\to 2\chi}$ and numerically for  $\gamma_{2\psi\to 2\chi,\rm off}$, where the analytic relic density   from the decay channel can be written as
\begin{align}\label{eq:omegad}
\Omega_{\phi\to 2\chi} h^2\approx 0.34  \left(\frac{y_\psi}{0.1}\right)^3 \left(\frac{y_\chi}{ 10^{-9}}\right)^2
\end{align}
in the limit of  a generically weak coupling $y_\psi\lesssim 1$ and the Boltzmann approximation. 

Note that $\Omega_{\phi\to 2\chi}\propto y_\psi^3$ while $\gamma_{\phi\to 2\chi}\propto y_\psi^2$. The additional power dependence on the thermal coupling $y_\psi$ comes from the fact that the freeze-in DM production is IR dominated, and both the   decay $\phi\to \bar\chi+\chi$ and the annihilation $\bar\psi+\psi\to \bar\chi+\chi$ are kinematically forbidden at  zero temperature,  making the yield $Y_\chi$  depend  on the inverse threshold temperature and the heavy DM mass scale.  It can  then be  found   that both $\Omega_{\phi\to 2\chi}$  and $\Omega_{2\phi\to 2\chi,\rm off}$ are basically independent of  the DM mass.
However, for the vacuum mass condition $m_\psi>m_\chi$, the annihilation $2\psi\to 2\chi$ is opened at zero temperature and the yield  $Y_\chi$ would not have the simple  $1/m_\chi$ dependence.  This is the case for the sub-MeV or lighter DM production from the nonrelativistic electron-positron annihilation~\cite{Dvorkin:2019zdi,Chang:2019xva}.

Independent of the DM mass, the ratio of $\Omega_{2\psi \to 2\chi,\rm off} $ to $\Omega_{\phi\to 2\chi}$ can then be simply estimated by the thermal coupling $y_\psi$, which is shown in the left panel of Fig.~\ref{fg:DMrelic}. Approximately, we find that the ratio in the Boltzmann statistics can be   fitted as 
\begin{align}\label{eq:OmegaR}
	\frac{\Omega_{2\psi \to 2\chi,\rm off} }{\Omega_{\phi\to 2\chi} }\approx 0.8 y_\psi + 1.8 y_\psi^{-1}+0.5,
\end{align}
while the ratio in the full quantum statistics is approximately a factor of 2 smaller than in   the Boltzmann statistics.
For smaller thermal coupling $y_\psi$, the ratio basically scales as $1/y_\psi$. Taking $y_\psi=10^{-3}$ for example, we can see that the DM relic density produced through the off-shell scattering channel is a factor of 1800 larger than that through the forbidden decay, however the ratio decreases below $\mathcal{O}(10)$ for  an electroweak gauge coupling.   When   $y_\psi=1$, the ratio in Eq.~\eqref{eq:OmegaR} gives $\Omega_{2\psi \to 2\chi,\rm off}/\Omega_{\phi\to 2\chi}\approx 3.1$. 
It indicates that the contribution to the DM relic density from the forbidden decay becomes significant when the thermal coupling is large. Typically, we expect a   portion of 5$\%$-24$\%$ (10$\%$-39$\%$) from the forbidden decay in the  weak-coupling regime $0.1\lesssim y_\psi\lesssim1$ when the Boltzmann (quantum) statistics is used. 

We can also see from  the right panel of Fig.~\ref{fg:DMrelic} the correlation between the  nonthermal DM coupling $y_\chi$ and the thermal coupling $y_\psi$ when     the observed DM relic density $\Omega_{\rm obs}h^2=0.12$~\cite{Planck:2018vyg} is accounted for by the sum of  the forbidden decay and  scattering channels. Besides, we also show the correlation when the DM relic density is only generated by the forbidden decay.   
As seen from the right panel of Fig.~\ref{fg:DMrelic}, when compared to the purely forbidden decay channel, the strong scattering contribution with a small thermal coupling  opens up the parameter space of the   DM coupling towards  smaller values. We can also observe that the Boltzmann approximation does not lead to large discrepancies from the full quantum statistics.

\textbf{Discussion}.---While we   consider a simple scalar mediator here,   the similar pattern between the scattering and the forbidden decay can also be expected in other light mediators~\cite{Li:2023ewv}. For instance,  a  fermion DM coupling  to the SM via a vector mediator (the photon) has been considered in Refs.~\cite{Dvorkin:2019zdi,Chang:2019xva}.  It was  demonstrated that the plasmon decay is the dominant channel for sub-MeV DM production. This can in fact be explained by Fig.~\ref{fg:resonant}. The effective  photon  thermal mass can be estimated by a  correction factor  $\kappa\sim \sqrt{\alpha_{\rm EW}}\sim \mathcal{O}(0.1)$ in the thermal plasma. For a sub-MeV DM the threshold temperature  can reach   $T_c\sim 10^{-3}$~GeV and   the sudden drop of the decay curve in Fig.~\ref{fg:resonant} can be postponed until $m_\chi/T\simeq \mathcal{O}(0.1)$. When the temperature is above the electron mass $m_e$, the contributions from the plasmon decay   and  the  electron-positron pair production are comparable. However, the  pair production  becomes Boltzmann suppressed when $T_c<T<m_e$, and the scattering curve in Fig.~\ref{fg:resonant} would exhibit the drop prior to that in the decay curve. In this case, the scattering contribution is suppressed in  the   history of the production and the plasmon decay becomes the dominant channel. For much heavier DM, however, the production from the nonrelativistic electron-positron annihilation is kinematically forbidden. In this case, the collision rates from the  plasmon decay and the  electron-positron scattering are expected to have   similar patterns shown in Fig.~\ref{fg:resonant}.

The relation between the  scattering and the forbidden decay can also have    important consequences  in the scenarios of  right-handed neutrino    DM production~\cite{Besak:2012qm,Drewes:2015eoa}. When the right-handed neutrino   $N_R$ is nonthermally produced by  some forbidden decay   at  higher temperatures, the contribution from the related scattering sensitively depends on the mediator connecting the SM and $N_R$. If the mediator is  the SM Higgs, which carries  $\kappa\approx 0.4$~\cite{Cline:1993bd} from   gauge   and top Yukawa interactions, both the scattering and forbidden decay could contribute to the  $N_R$ production comparably~\cite{Giudice:2003jh}.
However, if a right-handed neutrino   only couples  to  a  thermal scalar singlet that has a weaker connection to the SM plasma,   it can be be inferred from Fig.~\ref{fg:resonant} and Eq.~\eqref{eq:OmegaR} that  the contribution    from the forbidden decay  would be much smaller than the scattering channel.

The out-of-equilibrium scattering production associated with   the forbidden decay can also  modify  the thermally induced generation of lepton asymmetries in the early universe~\cite{Giudice:2003jh,Hambye:2016sby,Hambye:2017elz,Li:2020ner,Li:2021tlv}. If some forbidden decay is opened to produce  $N_R$ in a $CP$-violating way,  the $CP$ asymmetry stored in $N_R$ can be transferred to the SM one, which in the active sphaleron epoch is partially converted into the baryon asymmetry. If the mediator carries a thermal parameter $\kappa\ll 1$,  the scattering channel can  be readily stronger than the forbidden decay.

\textbf{Conclusion}.---
We have illustrated a close relation  between the forbidden decay and the associated  scattering     at finite temperatures.  Instead of carrying higher- or lower-order weak coupling constants, the two-body scattering hinted from  the forbidden  decay diagram  can carry the same order of coupling constants. For a light scalar mediator discussed in this paper,  the close relation can be simply described by the dominant thermal coupling between the plasma and the mediator.   Such a simple relation allows us to estimate the relative contribution to the DM relic density from the scattering and the forbidden decay. We found that   the forbidden decay becomes important and can contribute to the total DM relic density at 5$\%$-24$\%$ (10$\%$-39$\%$) for a thermal interaction   in the weak-coupling regime and in the Boltzmann (quantum) statistics, but the scattering effect increases at the speed of inverse thermal coupling when the connection between the plasma and the mediator is much weaker.

 The author thanks  Oleg Lebedev,  Katelin Schutz and Xun-Jie Xu  for valuable discussions. This work is supported in part by the National Natural Science Foundation of China under grant No. 12141501.

\bibliographystyle{JHEP}
\bibliography{Refs}

\end{document}